# Continuously Differentiable Analytical Models for Implicit Control within Power Flow

Aayushya Agarwal[1], *Graduate Student Member, IEEE*, Amritanshu Pandey[1], *Graduate Student Member*, *IEEE*, Marko Jereminov[1], *Graduate Student Member*, *IEEE*, and Larry Pileggi[1], *Fellow, IEEE*

*Abstract*— Achieving robust and scalable convergence for simulation of realistic power flow cases can be challenging. One specific issue relates to the disconnected solution space that is created by the use of piecewise-discontinuous models of power grid devices that perform control mechanisms. These models are generally resolved by outer iteration loops around power flow, which can result in solution oscillations, increased iteration count, divergence or even convergence to a solution in an unstable operational region. This paper introduces a continuously differentiable model for device control mechanisms that is incorporated within the power flow formulation. To ensure robust power flow convergence properties, recently introduced homotopy methods are extended to include these continuous models. The scalability and efficacy of the proposed formulation is demonstrated on several large-scale test cases that represent the US Eastern Interconnect network, the Synthetic USA, and the Nigerian grid.

*Index Terms*--continuous analytical models, divergence, homotopy methods, piecewise discontinuous models, power flow, split-circuit formulation

## I. INTRODUCTION

In order to effectively study the transmission and distribution of power, it is necessary to simulate the power grid operation under various conditions and scenarios. Although a robust power grid simulator must be scalable and robustly converge to a stable solution for any examined test case, traditional formulations can suffer from convergence issues [1].

Power flow is an analysis that solves for a steady state operating point of the interconnected grid. To simulate the realistic grid behavior, additional constraints that represent the control loops of controllable grid devices have to be incorporated within the simulation framework. These control loops are typically modeled as piecewise discontinuous functions, thereby making the solution space not only highly non-linear, but also discontinuous [2]. Specifically, enabling reactive power limits of generators through PV/PQ switching, enabling reactive power limits of discrete switched shunts and transformer taps (tap limit), and modeling distributed slack with active power limits (mimics automatic generation control (AGC) in power flow), are all modeled as piecewise discontinuous functions. Most importantly, the use of piecewise discontinuous models prevents the explicit use of numerical methods such as Newton Raphson (N-R) to solve the corresponding power flow problem.

Traditional power flow formulation applies N-R for solving the continuous models, then uses outer loops to switch through segments of piecewise models for the aforementioned control devices. It can be shown [3] that the application of these fixed-point iteration outer loops to the discontinuous space in power flow is prone to oscillations, increased iteration count, divergence or even convergence to a solution in the unstable operational regions [4]-[5].

The problems associated with piecewise functions in power flow has been widely recognized, dating back to [1]. In [1], the authors reviewed the load flow algorithms and described two methods to suppress oscillations created by the discontinuous PV model. The first method, most commonly used today, controls oscillations by fixing the PV bus as a PQ after a set number of iterations, but can result in convergence to unstable operating regions for generators. The authors in [3] presented multiple testcases that diverged, oscillated, or resulted in unstable operation conditions. Furthermore, the supplementary heuristics that were claimed to prevent oscillations, introduced in [3], significantly increased the iteration count for the simulation, and were not shown to scale well with system size. The second method introduced by [1] proposed the use of an error correction term in the decoupled power flow to limit a generator's reactive power. This technique was extended to AC power flow [6], however it was found to result in excess iterations and oscillations for reactive limits of generators. This technique with error correction term is also used for modeling distributed slack in the system (to mimic AGC) [7] via participation factors and is solved implicitly within the N-R inner loop. However, when a generator violates its active power limits while participating within this scheme, it must be resolved in the outer loop.

For a more robust and accurate simulation, continuous models for control functions have been previously proposed to eliminate discontinuities in the power flow formulation with these devices; however, most of these methods have been limited to reactive power control of generators through PV/PQ switching. The authors in [8] implemented a smooth PV model using a sigmoid function to represent the control of reactive power limits of a generator. With this continuously differentiable function, the authors were able to converge for

This work was supported in part by the Defense Advanced Research Projects Agency (DARPA) under award no. FA8750-17-1-0059 for RADICS program, and the National Science Foundation (NSF) under contract no. ECCS-1800812.

[1]Authors are with the Department of Electrical and Computer Engineering, Carnegie Mellon University, Pittsburgh, PA 15213 USA (e-mail: {aayushya, mjeremin, amritanp, pileggi}@andrew.cmu.edu).

smaller test cases without encountering oscillations. However, the authors did not demonstrate the methodology for larger test cases and did not account for remote generators in the formulation. In general, continuous models for other control loops, such as discrete devices and distributed slack, have not yet been explored comprehensively, and there is no framework for robust and scalable power flow formulations that include such models.

Recent formulations of power flow in terms of equivalent circuit elements modeling the N-R linearized equations enable the use of techniques that are analogous to those that are used for circuit simulation of highly nonlinear electronic device models [9]. These techniques have enabled robust simulation convergence for large-scale electronic systems that contain continuous, analytical models of transistors, diodes, and other circuit elements with complex nonlinearities [10]-[12]. We propose to extend this equivalent circuit-based framework for power systems to model the control loop mechanisms implicitly within the power flow problem [13]. The main contribution of this paper is the introduction of continuous power flow models for control loop mechanisms corresponding to generator voltage control (PV/PQ switching), remote voltage controls, switched shunts, transformer taps and distributed slack power. Correspondingly, we will extend the recently developed circuit-based homotopy methods [14] to accommodate these models and ensure robust convergence for large scale systems, thereby obviating the need for outer loop iterations.

The rest of the paper is structured as follows. Section II provides a brief background on the equivalent I-V based formulation and corresponding homotopy methods, as well as background on some of the commonly used power flow models and highlights the challenges with these piecewise discontinues models. Section III introduces two general analytical continuous models that are described in Section IV. In Section V, the continuous models are validated by comparing the solutions against the those obtained from existing tools for 75k+ nodes Eastern Interconnection test cases and other large synthetic test cases [15]. We present results for an actual Nigerian grid planning case that is prone to oscillations with conventional models but converges to a stable operating solution with the proposed analytical models. Lastly, we show that modeling distributed slack power in terms of continuous function eliminates the need for an outer loop.

## II. BACKGROUND

### A. Current-Voltage Based Power Flow Framework

Modeling of the transmission grid system has been traditionally handled as a power flow problem expressed in terms of a positive sequence formulation using real and reactive power (P and Q) state variables. Current-Voltage (I-V) based formulations, namely those based on Current Injection Method [16], have been primarily limited to solving distribution grid systems due to challenges with incorporating PV buses [17].

Recent formulations of power flow in terms of (I-V) state variables have been used to map the corresponding Newton-Raphson (N-R) nonlinear solution to an equivalent circuit model [9]. The resulting equivalent I-V formulation that corresponds to each linearization step of the N-R process can now be solved using circuit simulation-based methods that constrain the individual circuit element terms [18]. This formulation also enables the use of circuit-element based homotopy methods to robustly solve large-scale transmission systems, or any system that includes numerous PV buses. Importantly, it is by building the power grid equivalent circuit representation that we derive heuristics to exploit the physical characteristics of the simulation problem.

As an example of this I-V formulation, a generator with an automatic voltage regulator (AVR) is modeled as a PV bus in power flow and aims to control the bus voltage magnitude to a set value as given by (1). In an I-V based formulation, the PV bus is modeled as real and imaginary non-linear currents ($I_{RG}$ and $I_{IG}$), derived from first principles, as shown in (2)-(3). It is important to note that in addition to real and imaginary voltage variables, $V_{RG}$ and $V_{IG}$, the reactive power, $Q_G$, is added as a variable in order to control the voltage magnitude to a given set point from (1). To facilitate the use of N-R, the PV bus, as well as the other nonlinear devices, are linearized, and each term can be shown to map into equivalent circuit elements, the values of which are updated at each iteration. For instance, the linearized PV bus equivalent circuit shown in Figure 1 is based on the derivations in [16].

$$V_{set}^2 = V_{RG}^2 + V_{IG}^2 \tag{1}$$

$$I_{RG} = \frac{P_G V_{RG} + Q_G V_{IG}}{V_{RG}^2 + V_{IG}^2} \tag{2}$$

$$I_{IG} = \frac{P_G V_{IG} - Q_G V_{RG}}{V_{RG}^2 + V_{IG}^2} \tag{3}$$

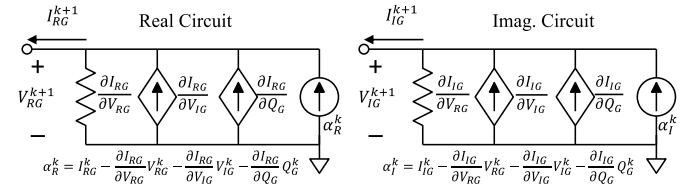

Figure 1 Linearized power flow split circuit of a PV generator.

### B. PV/PQ switching

In addition to having fixed inputs for active power, $P_G$, and voltage, $V_{set}$, a generator with an AVR has the reactive power limits, $Q_{max}$ and $Q_{min}$. Once a generator's reactive power, $Q_G$, exceeds its limits, it can no longer control the voltage, which is expressed in existing frameworks as a switch from a PV to PQ bus model, thus allowing the bus voltage to move from the set point. In an I-V based formulation, this results in replacing (1) by (4) or (5).

$$\text{if } Q_G \geq Q_{max}, \text{ then } Q_G = Q_{max} \tag{4}$$
$$\text{if } Q_G \leq Q_{min}, \text{ then } Q_G = Q_{min} \tag{5}$$

Importantly, this switching between PV and PQ models represents a piecewise discontinuous function within a power flow simulation. Hence, after each NR convergence, the outer loop must check if the PV bus has violated its limits and if so, back-off $Q_G$ to $Q_{max}$ or $Q_{max}$ and recast the PV to a PQ model.

Additionally, the generator must operate within a stable region, derived from the capability curves [19], as shown in Figure 2. Overall, the $dV_G/dQ_G$ sensitivity of the generator must be positive in order to remain in the stable region.

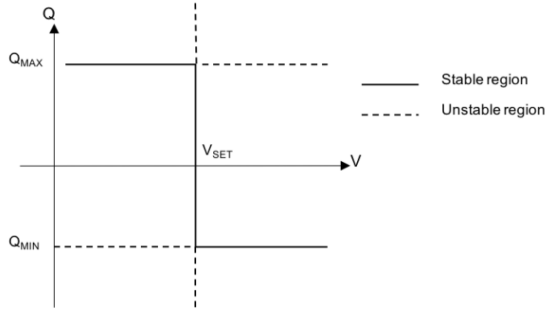

*Figure 2 Generator QV curve defining the stable and unstable operation regions.*

Multiple generators may be required to control the voltage of a single remote bus using a remote bus participation scheme. This is modeled by distributing their reactive powers based on a pre-defined vector of participation factors. Namely, each generator within the group contributes a certain percentage of the reactive power required to control the voltage, while remaining within its reactive power limits. When a generator within the group violates its limits and must switch to a PQ model, the other generators controlling the remote bus contribute the resultant slack reactive power by re-adjusting their participation factors. In existing power flow simulators, the reallocation of participation factors is a piecewise function performed in the outer loop, which creates new discontinuities within the model.

*C. Issues with Piecewise Models*

Using piecewise continuous models that are not first-derivative continuous within nonlinear system solvers relies on stringent conditions for convergence [20]. Numerical methods such N-R, require first derivative continuity, and therefore cannot be applied to such models. To accommodate piecewise models, power flow methods have traditionally handled them with an outer iteration loop. These outer loop iterations are applied to the control mechanisms such as PV/PQ switching, switched shunt and transformer tap control, as well as distributed slack control. But the discontinuous models split the solution space into disjoint regions, which can result in divergence and oscillatory behavior when these outer loop iterations are applied.

Piecewise linear models for active elements (e.g. transistors) have been used in circuit simulation. The authors in [20] demonstrated that linear piecewise functions are guaranteed to converge as long as no more than one model switches from one segment to an adjacent segment during any iteration step. This is an extremely difficult constraint for power flow, since during simulation of larger systems, many piecewise models exceed their limits during each inner loop of N-R simulation. Furthermore, allowing one segment change at a time would significantly increase the iteration count and result in different solutions based on the order of models that are being changed. We illustrate this with our results for the Eastern Interconnect test case, shown in Table 1. The simulations are based on using conventional discontinuous models that result in multiple generators violating their limit, thereby requiring PV-PQ switching. As shown in the results, the simulation problem converges to different solutions depending on the order of generators being switched. The table compares results based on starting with the smallest generator versus the largest generator.

TABLE 1 RESULTS OF EASTERN INTERCONNECT TESTCASE WITH DIFFERENT ORDER OF GENERATOR PV-PQ CONVERSIONS

| Gen. Conversion order | $V_{MAX}$ [pu] | $V_{MIN}$ [pu] | $\Theta_{MAX}$ [deg] | $\Theta_{MIN}$ [deg] |
|---|---|---|---|---|
| Largest first | 1.484 | 0.775 | 177.74 | -152.51 |
| Smallest first | 1.477 | 0.774 | 178.50 | -151.62 |

Piecewise models are also known to result in oscillations, particularly when the final operating point approaches the point of model discontinuity. There are many methods to suppress these oscillations, as outlined by [1]. The most common is to set a maximum limit on the number of conversions between PV and PQ models, and once that limit is met, fix the bus as a PQ. Once the generator is fixed as a PQ model, it no longer is able to control the voltage. This often results in generators operating in unstable regions of their Q-V curve (Figure 2), as shown for the Nigerian 3500 testcase.

The Nigerian 3500 testcase represents a real 550 node system grid with net load of 3500 MW that is found to be difficult to converge because many generators are operating close to their lower reactive power limit. When using a discontinuous model, this testcase oscillates between the PV and PQ segments until the generator is fixed as PQ. Figure 3 shows a plot of the number of generators that convert from PV to PQ and vice versa during each N-R outer loop when solving the case from a flat-start. Evidently, there are several oscillations that are suppressed by fixing the generator as a PQ segment. When this happens, generators no longer control their voltages and operate in unstable regions; namely, they supply the minimum reactive power and yet the voltage is lower than the set point. Most importantly, by fixing the generator model after several oscillations, power flow solvers are further constraining the solution space that may result in divergence if a feasible operating point in the over-constrained solution space doesn't exist.

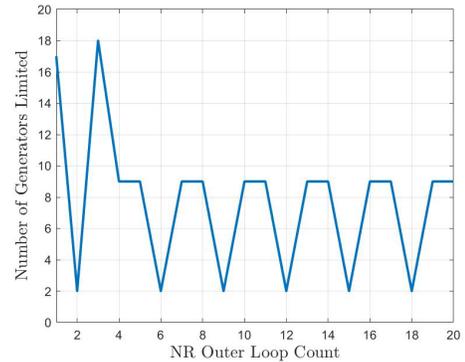

*Figure 3 Oscillations of PV/PQ conversions in Nigerian 3500MW testcase*

*D. Homotopy Methods*

It is possible to use highly nonlinear smooth models with continuous first-derivatives to approximate piecewise functions; however, they can make transmission systems increasingly difficult to simulate. Circuit simulation based homotopy methods can be developed to accommodate such nonlinearities and enable robust convergence to a solution.

The homotopy method aims to solve a series of easier sub-problems that sequentially lead to solving the original power flow problem. Namely, it traces a path in the solution space, which can be mathematically described as

$$\mathcal{H}(x, \lambda) = (1-\lambda)\mathcal{F}(x) + \lambda\mathcal{G}(x), \forall \lambda \in [0,1] \quad (6)$$

where, $\lambda$ is a homotopy factor that starts at a value of $\lambda_{init}$, which thereby represents a trivial problem, $\mathcal{G}(x)$, then is progressively reduced to a value of 0 to represent the original problem, $\mathcal{F}(x)$. In general, a homotopy factor $\lambda \in [0, \lambda_{init}]$ is embedded into the power flow models, thereby relaxing the models to a trivial representation at $\lambda = \lambda_{init}$ which corresponds to a decrease in the strength of the non-linearities. With each subsequent homotopy factor, we solve the network of equations and decrement the homotopy factor if the modified network has converged. We summarize what a homotopy flow would look like for a power flow problem in Figure 4.

$$y = \frac{y_{max} - y_{min}}{1 + \exp(S(x - x_{set}))} + y_{min.} \quad (7)$$

where, $x_{set}$ is the nominal set point of the variable $x$, while $y_{min}$ and $y_{max}$ represent the limits of the variable $y$. Furthermore, $S$ represents a smoothing factor that determines the steepness of the sigmoid curve, as shown in Figure 5. It is important to note that a larger smoothing factor mimics the behavior of a piecewise discontinuous model.

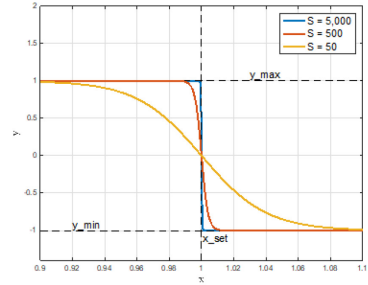

*Figure 5 Continuous sigmoid curve with minimum and maximum limits for various smoothing factor.*

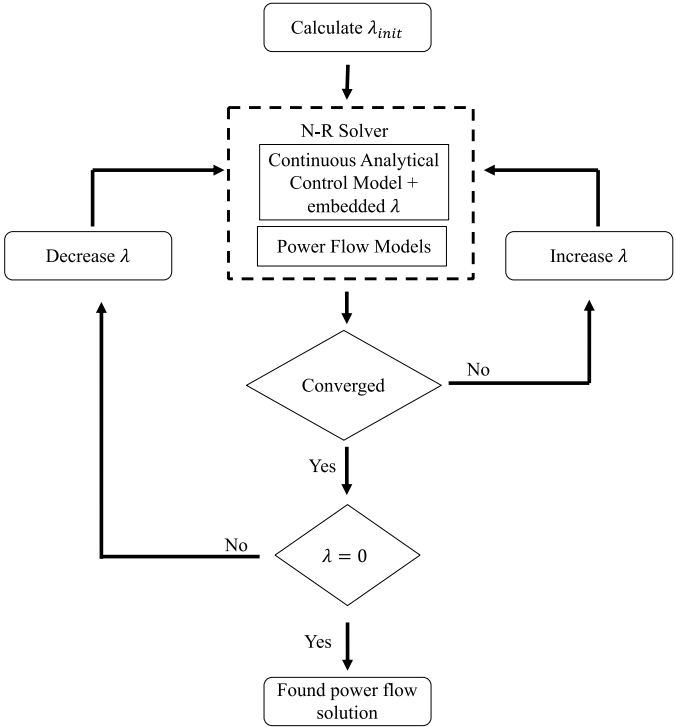

*Figure 4 General Flow Diagram for Homotopy Method*

Using the equivalent circuit formulation for power flow, Tx-stepping was recently developed as a homotopy method for ensuring convergence of large-scale transmission system simulation from arbitrary initial conditions [14]. This method can guarantee convergence under certain reasonable assumptions.

Importantly, the Tx-stepping homotopy factor $\lambda$ is embedded within the linearized (during N-R) network elements. Therefore, the initial problem, $\mathcal{G}(x)$, represents a virtually shorted network, whose solution is feasible, and governed by the generator voltages and slack bus angles. Next, the problem is solved by sequentially reducing $\lambda$ to zero, while using the previous solution as an initial condition and resolving the power flow until the network returns to the original state.

### III. CONTINUOUS ANALYTICAL MODELS

In this section we present the generic continuously differentiable functions that represent the basic building blocks for our continuous analytical power flow control models.

#### A. Sigmoid Model

We introduce the following sigmoid curve that is further used to model the machine physical limits:

#### B. Participation Model

Machines in the power grid are often required to coordinate dispatching resources with each other. For example, a group of generators may coordinate the amount of reactive power they must dispatch to control the voltage at a remote bus. These machines also have physical limits on the amount they can dispatch. In general, each machine contributes a ratio of reactive power, known as the participation factor, $\kappa$.

In existing power flow tools, the participation factor of a machine is pre-defined and used implicitly within the N-R iterations. However, the physical limits are monitored in the outer loop and if violated, modeled as piecewise discontinuous functions. Furthermore, the participation factors are updated, and the inner loop of N-R is re-run until a solution is obtained with control equipment output within its limits.

Herein, we introduce a piecewise continuous function that establishes the participation factor while upholding the physical limits implicitly, thereby enabling its direct incorporation within the inner loop of N-R. Equation (8) illustrates the model as five segments: three linear segments and two quadratic patching functions. The small quadratic patches allow the function to remain continuously differentiable since the first-order derivatives of the piecewise functions match at the intersecting points.

$$y = \begin{cases} y_{min}, & \text{Region 1} \\ a_{min}x^2 + b_{min}x + c_{min}, & \text{Region 2} \\ \kappa x, & \text{Region 3} \\ a_{max}x^2 + b_{max}x + c_{max}, & \text{Region 4} \\ y_{max}, & \text{Region 5} \end{cases} \quad (8)$$

The variable, $x$ is common to all the machines within the same group and allows each machine to coordinate its dispatch with the other machines within the group. The $y$ variable represents the dispatch output for an individual machine. Each machine cooperating in the dispatch will apply its own participation function with a common variable, $x$ shared within the group. The participation model includes each machine's dispatch limits, as given by regions 1 and 5. Region 3 is the linear region which controls the amount a machine will dispatch, and the slope is equal to the participation factor, $\kappa$. Most importantly, when one machine hits its limits, i.e. settles into regions 1 or 5, the other machines do not have to be recalculated because their dispatch is controlled with respect to the common variable.

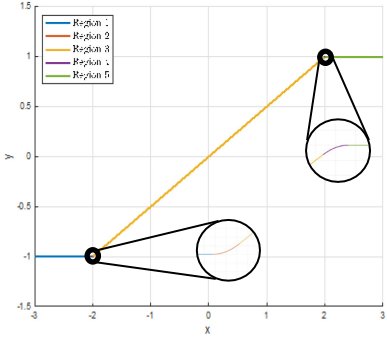

Figure 6 Continuous patching function with minimum and maximum limit.

## IV. Power Flow Models And Homotopy Methods

Armed with the continuously differentiable models from the previous section, we now seek to build the control models that we will incorporate within our power flow implementation.

### A. PV Model

The equations for the real and imaginary currents of the PV model remain as shown in (2)-(3). An extra equation is required to model the control of reactive power and the voltage magnitude while inherently considering reactive power limits. The sigmoid curve is a suitable choice with the y axis representing $Q_G$ and the dependent variable being the voltage magnitude $V_{set}$ at the bus. The continuous function from (7), can be then written as:

$$Q_G = \frac{Q_{max} - Q_{min}}{1 + \exp\left(S\left[\sqrt{V_{RG}^2 + V_{IG}^2} - V_{set}\right]\right)} + Q_{min} \quad (9)$$

The continuous PV model mimics the behavior of a discontinuous model implemented in the outer loop when the smoothing factor is large (a large value of 5000 corresponds to a maximum error of 0.4%). It mimics the operation in both the "voltage set" mode given by (1) and the "set reactive power" mode given by (4) and (5), but is a continuous and differentiable function. When the generator reactive power is within its bounds, the sigmoid curve controls the magnitude of bus voltage to $V_{set}$. However, if the reactive power hits its limits, the reactive power saturates while the voltage magnitude can deviate from its set-point. Importantly, the sigmoid restricts a PV bus model to converge to a stable operating point only, unlike the conventional piecewise model.

Nonetheless, a sigmoid curve with a large smoothing factor is highly nonlinear and can encounter convergence difficulties for large systems. The derivative of the sigmoid curve far away from the center is nearly zero, meaning it could take significant number of N-R steps to reach the center. Also, the steep edges of the function near the limits have large derivatives causing simulation convergence issues. To provide convergence robustness we have developed and applied the following two homotopy methods.

#### 1) Relaxing generator smoothing

The first approach embeds a homotopy factor into the sigmoid function to control the steepness, as shown in (10). The steepness, $(S - \lambda_S)$, determines how closely the sigmoid function mimics the piecewise discontinuous function. By reducing the steepness, the function departs from the approximate behavior of the discontinuous model while reducing the severity of the non-linearities. In this method, we first obtain an initial value of $\lambda_S$, $\lambda_{S,init}$, for which the sigmoid function does not include large non-linearities. For the analysis and results shown in the paper, we set $\lambda_{S,init}$ such that the initial steepness, $(S - \lambda_{S,init}) = 100$. After solving the initial problem, we decrease each generator's homotopy factor, $\lambda_S$ toward zero until the original problem is solved following the flow in Figure 4.

$$Q_G = \frac{Q_{max} - Q_{min}}{1 + \exp\left((S - \lambda_S)\left[\sqrt{V_{RG}^2 + V_{IG}^2} - V_{set}\right]\right)} + Q_{min} \quad (10)$$

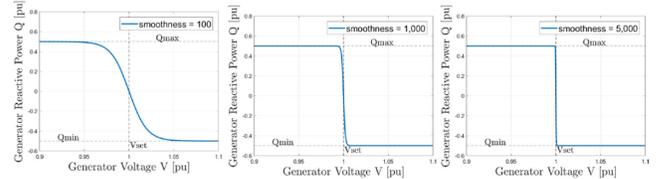

Figure 7 Effect of homotopy for generators smoothing factor relaxation.

#### 2) Linear relaxation of reactive power limits

For this approach, we embed a homotopy parameter $\lambda_G$ into the continuous generator model:

$$Q_G = \frac{\lambda_G(Q_{MAX} - Q_{MIN})}{1 + exp(S\left[\sqrt{V_{RG}^2 + V_{IG}^2} - V_{set}\right])} + \lambda_G Q_{MIN} \quad (11)$$

As with the previous homotopy method, the value of $\lambda_G$ that results in a trivial solution must be used initially. This is achieved by calculating $\lambda_G^{init}$ by solving the inner loop of the power flow problem with unbounded reactive power limits for the generator and choosing its value as:

$$\lambda_G^{init} = \begin{cases} \dfrac{Q_G}{Q_{MAX}}, & \text{if } Q_G > Q_{MAX} \\ \dfrac{Q_G}{Q_{MIN}}, & \text{if } Q_G < Q_{MIN} \\ 1, & \text{otherwise} \end{cases} \quad (12)$$

Once $\lambda_G^{init}$ is obtained, we incrementally vary the parameter $\lambda_G$ as shown in Figure 4. The range of generator convergence parameter $\lambda_G$ is given by $[1, \infty)$. Figure 8 graphically demonstrates the introduced methodology.

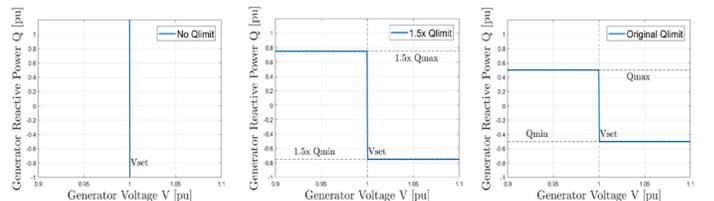

Figure 8 Homotopy for generators reactive power limits relaxation.

### B. Remote Voltage Control

Numerous power grid elements such as generators, FACTS devices, transformers, shunts, etc., can control a voltage magnitude at a given bus in the system. Moreover, they can control the voltage magnitude at either their own bus ($\mathcal{O}$) or a remote bus ($\mathcal{R}$) with a single or a set of devices. Here, we define a continuous model for the remote voltage control.

When a set of generators control the voltage of single remote node, the following conditions must be met:
i. Remote generators participating in controlling a remote bus must deliver reactive power with respect to a participation factor and must stay within its reactive power limits.
ii. If all the remote generators have hit their reactive power limits, the voltage of the bus may be allowed to change within a generator's stable operating region.

To meet the first condition, we apply the participation model with the limits representing an individual generator's reactive power limits as

$$Q_G = \begin{cases} Q_{G,MIN}, & \text{Region 1} \\ a_{min}Q_{REQ}^2 + b_{min}Q_{REQ} + c_{min}, & \text{Region 2} \\ \kappa\, Q_{REQ}, & \text{Region 3} \\ a_{max}Q_{REQ}^2 + b_{max}Q_{REQ} + c_{max}, & \text{Region 4} \\ Q_{G,MAX}, & \text{Region 5} \end{cases} \quad (13)$$

The reference variable, $Q_{REQ}$, is the total reactive power required by a remote bus to the set voltage point. Furthermore, when none of the remote generators in the group have hit their reactive power limits, $Q_{REQ}$ is equal to the sum of the reactive power produced by each generator.

To meet the second condition, we utilize the sigmoid model from (7). This model describes the relationship between the requested reactive power of the remote bus to the voltage at that bus. The dependent variable, $Q_{REQ}$ is the requested reactive power and the x-axis is the voltage magnitude $V_{set}$ of the remote bus given by (14). The maximum and minimum limits for reactive power are $Q_{REQ,max}$, $Q_{REQ,min}$, respectively and are given in by the sum of the reactive power limits of each participating generator, $Q_{i,max}$ and $Q_{i,min}$ (15)-(16).

$$Q_{REQ} = \frac{Q_{REQ,max} - Q_{REQ,min}}{1 + \lambda_S \exp\left(\sqrt{V_R^2 + V_I^2} - V_{set}\right)} + Q_{REQ,min} \quad (14)$$

$$Q_{REQ,max} = \sum Q_{i,max} \quad (15)$$

$$Q_{REQ,min} = \sum Q_{i,min} \quad (16)$$

### C. Discrete Control Devices

Discrete control devices, such as switched shunts and transformer taps, regulate their output in discrete steps while maintaining their output within operational limits. Typically, discrete control requires an outer loop to facilitate changes in the output. We can approximate the possible discrete values of the control device by mapping it to a continuous space using a sigmoid curve that represents the possible values of the controlled output within the operational limits. In order to model the realistic behavior, the continuous value must be snapped to the closest discrete value to obtain the final solution.

In practice, it is rare that this snapping to the closest discrete value could cause the simulation to diverge. However, theoretically it is possible for two reasons:
i. The modified device state due to the change from its continuous state to discrete state could result in an infeasible network.
ii. The set of non-linear equations representing the modified system state may diverge when initialized from prior solution.

In case the system is infeasible due to the snapping action, we can utilize an optimization-based method for evaluating power flow feasibility [21]. The approach in [21] is incorporated within the same power flow simulation engine and can be used to identify the system infeasibility and accordingly adjust the discrete element values such that the system is feasible. For cases where the divergence is due to a poor initial condition for the snapped system state, continuation methods can be used to gradually modify the discrete elements parameters from their continuous value to discrete value until convergence is achieved. This model for discrete control devices allows us to effectively model discrete switched shunts and transformer taps.

A discrete switched shunt is responsible for producing reactive power within operational limits to maintain the voltage in a given range. We model the switched shunt as a continuous local generator with zero active power. A continuous sigmoid curve approximates the possible values of reactive power supplied within the reactive power limits. Then to snap the reactive power to the closest discrete step, we use the aforementioned techniques. To overcome the nonlinearities introduced by the sigmoid function, we apply the homotopy methods that were described for the continuous PV model.

Transformers with controllable taps are another form of discrete control devices that adjust their tap ratio based on the voltage at the controlling buses. Within the limits of a transformer tap, the voltage at the bus is controlled to a set value. If the voltage on the primary side is lower than the set voltage, the transformer taps on the primary side increase. On the other hand, if the voltage is higher, then the primary taps decrease to bring the voltage back to the set point. If the transformer tap is unable to control the voltage, it sets its turns ratio to a constant value (either a maximum or minimum). Similarly, the secondary side has the reverse relation. This relationship is modeled by the sigmoid function from Figure 9.

To mimic the control of the turns ratio $tr$, we use the continuous sigmoid curve from Figure 5. The equation for the transformer taps on the primary side is given by (17), and the equation for the secondary side is given by (18), where $tr_{max}$ and $tr_{min}$ refer to the maximum and minimum transformer tap ratios, respectively.

$$tr = \frac{tr_{max} - tr_{min}}{1 + \exp\left(\lambda_S\left(\sqrt{V_R^2 + V_I^2} - V_{SET}\right)\right)} + tr_{min} \quad (17)$$

$$tr = \frac{tr_{min} - tr_{max}}{1 + \exp\left(\lambda_S\left(\sqrt{V_R^2 + V_I^2} - V_{SET}\right)\right)} + tr_{max} \quad (18)$$

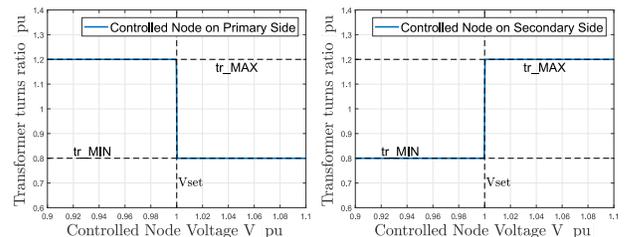

*Figure 9 Continuous transformer tap control with voltage for primary and secondary sides.*

## D. Distributed Slack Power

During a contingency, generators within an area may be required to change their $P_G$ to maintain the frequency at the nominal value. AGC coordinates participating generators to share excess active power, $\Delta P_G$, based on a ratio to the excess active power produced by the slack bus, $\kappa$. Moreover, generators have maximum and minimum limits ($P_{G,MAX}, P_{G,MIN}$) to the amount of active power they can produce, which are typically handled in an outer loop.

We apply the continuously differentiable participation formulation in Section III to model the control loop that distributes the slack power. Each generator's extra active power, $\Delta P_G$, is modeled with respect to the extra active power of the slack bus, $\Delta P_S$, with slope of the curve representing the generators participation factor $\kappa$. The limits of the participation model are the maximum and minimum excess active power a generator can dispatch, as given in (22)-(23).

$$\Delta P_G = \begin{cases} \Delta P_{G,MIN} + \lambda_P \Delta P_{MIN,extra}, & \text{Region 1} \\ a_{min}\Delta P_S^2 + b_{min}\Delta P_S + c_{min}, & \text{Region 2} \\ \kappa \Delta P_S, & \text{Region 3} \\ a_{max}\Delta P_S^2 + b_{max}\Delta P_S + c_{max}, & \text{Region 4} \\ \Delta P_{G,MAX} + \lambda_P \Delta P_{MAX,extra}, & \text{Region 5} \end{cases} \quad (19)$$

$$\Delta P_{G,MAX} = P_{G,MAX} - P_{G,INIT} \quad (20)$$
$$\Delta P_{G,MIN} = P_{G,MIN} - P_{G,INIT} \quad (21)$$
$$\Delta P_{MAX,extra} = \Delta P_G - \Delta P_{G,MAX} \quad (22)$$
$$\Delta P_{MIN,extra} = \Delta P_G - \Delta P_{G,MIN} \quad (23)$$

The excess powers must also be incorporated into the generator current terms and are shown as follows:

$$I_{RG} = \frac{(P_G + \Delta P_G)V_{RG} + Q_G V_{IG}}{(V_{RG})^2 + (V_{IG})^2} \quad (24)$$

$$I_{IG} = \frac{(P_G + \Delta P_G)V_{IG} - Q_G V_{RG}}{(V_{RG})^2 + (V_{IG})^2} \quad (25)$$

Lastly, to ensure the system is determined, the excess slack bus has to be mapped to an equation relating to its slack currents and voltages, producing the final expression:

$$P_S + \Delta P_S = V_{S,R} I_{S,R} + V_{S,I} I_{S,I} \quad (26)$$

The strict active power limits can make converging to a final solution difficult. Therefore, we apply our homotopy approach to first solve a system with no active power limits, $\Delta P_{G,MAX}$ and $\Delta P_{G,MIN}$. This modifies the participation model to a linear plot, with the slope equal to the $\kappa$ for the entire space. The solution is captured and the values of $\Delta P_{MAX,extra}$ and $\Delta P_{MIN,extra}$ are determined from (20) and (21). In addition, the homotopy factor, $\lambda_P$, is set to a value of 1 for the initial case. Then to reach the original solution, $\lambda_P$ is iteratively reduced to a value zero, which represents the generators with their original active power limits.

## V. RESULTS

### A. Validating Continuous Models

The continuous power flow models representing the control loops were incorporated into our prototype simulator, SUGAR [14]. The continuous models were validated by comparing the results of various testcases with commercial power flow simulators to verify that the voltages were matched. The testcases were a collection of network topologies representing sections of the US interconnected network [15][22]. Each testcase using the continuous models was executed from an arbitrary initial condition with limits on reactive power with a smoothing factor value of 5000.

It is important to note that these continuous models were co-developed with the homotopy methods, and they improved the efficiency of convergence for the sample test cases as compared to the same power flow engine with piecewise models in an outer iteration loop. The results for the Eastern Interconnect testcase are shown in Table 3. As an initial condition, we solved the Eastern Interconnect system without any generator reactive power or transformer tap limits. Using discontinuous models in SUGAR, we found that the solution oscillated between PV and PQ models many times before converging, which increased the iteration count.

TABLE 2 VALIDATING CONTINUOUS MODELS RESULTS BY COMPARING AGAINST COMMERCIAL POWER FLOW SOLVERS

| Test Case | Continuous Models in SUGAR | | | | Commercial Power Flow Simulator | | | |
|---|---|---|---|---|---|---|---|---|
| | $V_{MAX}$ | $V_{MIN}$ | $\Theta_{MAX}$ | $\Theta_{MIN}$ | $V_{MAX}$ | $V_{MIN}$ | $\Theta_{MAX}$ | $\Theta_{MIN}$ |
| Case13659Pegase | 1.18 | 0.84 | 98.57 | -34.71 | 1.18 | 0.84 | 98.57 | -34.57 |
| Eastern Interconnect | 1.50 | 0.77 | 178.82 | -151.0 | 1.50 | 0.77 | 178.82 | -151.0 |
| Synthetic USA | 1.09 | 0.77 | 109.62 | -109.87 | 1.09 | 0.77 | 109.62 | -109.87 |

TABLE 3 NUMBER OF ITERATIONS TO CONVERGE EASTERN INTERCONNECT USING CONTINUOUS AND DISCONTINUOUS MODELS

| Iterations | Continuous Models in SUGAR | Discontinuous Models in SUGAR |
|---|---|---|
| Eastern Interconnect | 67 | 94 |

### B. Nigerian Testcase

Next, we examined testcases with known convergence issues. Specifically, the Nigerian 3500MW testcase is designed to represent parts of the Nigerian grid with 41 generators, many of which are operating close to their reactive power limits. As a result, discontinuous models require many PV/PQ switches to converge, thereby leading to a solution where 11 generators are operating in an unstable region. In contrast, the continuous models constrain the generators to operate in a stable region, such that with the application of the corresponding homotopy techniques, convergence to correct results was obtained. The results comparison is shown in Table 4. The distribution of generator operating points on their Q-V curve is illustrated in Figure 10.

TABLE 4 RESULTS OF NIGERIAN 3500MW TESTCASE USING CONTINUOUS AND DISCONTINUOUS MODELS IN SUGAR

| Nigerian 3500MW | $V_{MAX}$ | $V_{MIN}$ | $\Theta_{MAX}$ | $\Theta_{MIN}$ |
|---|---|---|---|---|
| Continuous Models in SUGAR | 1.03 | 0.43 | 83.26 | -47.17 |
| Discontinuous SUGAR | 2.08 | 0.66 | 0 | -100.1 |

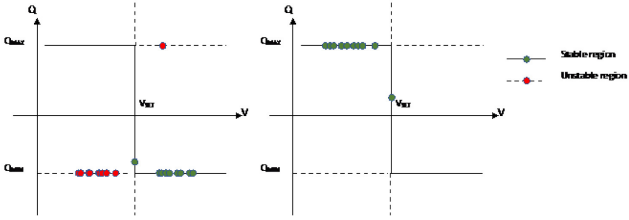

*Figure 10 Distribution of generator operating points on Q-V curve for Nigerian 35000MW Testcase based on a) using discontinuous models b) using continuous models.*

## C. Distributed Slack SAVWN Testcase

To validate the continuous distributed slack model we created a contingency on a sample 23-bus system, savnw. After running a base case with no contingency, we disconnected one generator on bus 211 and reported each generator's active power contribution with and without enabling distributed slack power. Table 5 illustrates that without distributing the slack power, the slack bus produces all of the excess active power, $\Delta P$. With the distributed slack power, the excess real power is distributed to each generator, which contributes a $\Delta P$ in proportion to a preset participation factor until the generators hit their limit. When a generator hits its limit, the other generators contribute extra active power based on the participation factor. In the testcase, every generator hit their maximum active power limit when they were sharing the excess slack power. More importantly, distributing the slack power was modeled continuously within the inner loop and did not require any outer loop controls.

TABLE 5 REAL POWER PRODUCED FOR EACH GENERATOR USING DISTRIBUTED SLACK POWER FOR SAVNW TESTCASE WITH AND WITHOUT CONTINGENCIES

| Generator ID | $P_{G,MAX}$ MW | $P_{G,MIN}$ [MW] | pf | Real Power Generation [MW] | | |
|---|---|---|---|---|---|---|
| | | | | Pre-contingency AGC-Disabled | Post-contingency AGC-Enabled | Post-contingency AGC-Disabled |
| 101 | 810 | 0 | 0.23 | 750 | 101 | 810 |
| 102 | 810 | 0 | 0.23 | 750 | 102 | 810 |
| 206 | 900 | 0 | 0.25 | 800 | 206 | 900 |
| 211* | 616 | 0 | 0.18 | 600 | 211* | 616 |
| 3011 | 900 | 0 | 0.08 | 257.74 | 3011 | 900 |
| 3018 | 117 | 0 | 0.03 | 100 | 3018 | 117 |

## VI. CONCLUSIONS

In this paper we introduced continuous analytical models for power flow control mechanisms that obviate the need of outer loop iterations. By using continuous sigmoid functions, along with patching functions and complementary homotopy methods, we demonstrated a robust approach for modeling control loops of AVR generators, remote generation, discrete switched shunts, transformer taps and distributed slack. The models were validated the models by comparison with piecewise discontinuous models for various large testcases and shown to provide a reliable solution for a testcase that would otherwise oscillate when simulated with outer loop iterations using a piecewise continuous model.